\documentclass[fleqn,10pt]{wlscirep}
\graphicspath{ {images/} }
\usepackage{subcaption}
\usepackage{multirow}
\title{Neural network interpolation of exchange-correlation functional}

\author[1, 2]{Alexander Ryabov}
\author[1]{Iskander Akhatov}
\author[1, *]{Petr Zhilyaev}

\affil[1]{Center for Design, Manufacturing and Materials, Skolkovo Institute of Science and Technology, Bolshoy Boulevard 30, bld. 1, Moscow, 143026, Russia}

\affil[2]{Moscow Institute of Physics and Technology (State University), Institutskiy per. 9, Dolgoprudny, Moscow Region 141700, Russia}

\affil[*]{p.zhilyaev@skoltech.ru}

\begin{abstract}
Density functional theory (DFT) is one of the most widely used tools to solve the many-body Schrodinger equation. The core uncertainty inside DFT theory is the exchange-correlation (XC) functional, the exact form of which is still unknown. Therefore, the essential part of DFT success is based on the progress in the development of XC approximations. Traditionally, they are built upon analytic solutions in low- and high-density limits and result from quantum Monte Carlo numerical calculations. However, there is no consistent and general scheme of XC interpolation and functional representation. Many different developed parametrizations mainly utilize a number of phenomenological rules to construct a specific XC functional. In contrast, the neural network (NN) approach can provide a general way to parametrize an XC functional without any \textit{a priori} knowledge of its functional form. In this work, we develop NN XC functionals and prove their applicability to 3-dimensional physical systems. We show that both the local density approximation (LDA) and generalized gradient approximation (GGA) are well reproduced by the NN approach. It is demonstrated that the local environment can be easily considered by changing only the number of neurons in the first layer of the NN. The developed NN XC functionals show good results when applied to systems that are not presented in the training/test data. The generalizability of the formulated NN XC framework leads us to believe that it could give superior results in comparison with traditional XC schemes provided training data from high-level theories such as the quantum Monte Carlo and post-Hartree-Fock methods.   
\end{abstract}

\begin{document}
\flushbottom
\maketitle

\thispagestyle{empty}

\section*{Introduction}

Since its initial formulation, density functional theory (DFT)~\cite{hohenberg1964inhomogeneous, kohn1965self} has become one of the main methods to solve the many-body Schrodinger equation. The source of the major ambiguity in DFT theory is the lack of a universal form of the exchange-correlation (XC) functional. Therefore, the success of DFT is mostly determined by the improvement of XC representations. The seminal Monte Carlo (MC) simulations of Ceperley and Adler~\cite{ceperley1980ground} have resulted in a number of practical local density approximations (LDAs)~\cite{vosko1980accurate, perdew1981self, perdew1992accurate}. The next major progress was produced by the introduction of the generalized gradient approximation (GGA), which takes into account local gradients of electron density~\cite{wang1991spin, perdew1992atoms, perdew1996generalized}. This greatly enhanced the ability of DFT to describe systems with inhomogeneous electron densities, such as molecules and surfaces. A number of other functionals have been suggested with various degrees of complexity, extending DFT to a broad class of material properties~\cite{mardirossian2017thirty}. The seeking for better functionals is still a very active area of research~\cite{lani2016adiabatic, maier2016new, mori2014derivative, mori2018exact}.

Despite the obvious success of the LDA and GGA, their construction is a highly complicated process that includes many heuristics steps. Even the XC functional form itself is based on some assumptions about the local nature of XC interactions. A large number of analytical conditions to fulfill~\cite{perdew1996generalized} and a commitment to the local functional form make it difficult to incorporate new numerical results from high-level quantum mechanics theories. For example, only numerical data from MC simulations of homogeneous electron gas are used to produce LDA~\cite{perdew1992accurate} and GGA~\cite{perdew1996generalized} functionals. Therefore, much of the progress in quantum MC~\cite{needs2009continuum, kolorenvc2011applications} and post-Hartree-Fock~\cite{cremer2011moller} methods is not adopted by XC approximations. One possible solution to overcome this problem is to use neural networks (NNs) for the interpolation of XC functionals. The NN provides a general way to represent any functional relationship~\cite{cybenko1989approximation} and a highly flexible framework to embody both analytical and numerical training data.

There have been a number of studies in which NNs have been used to describe the potential energy surface to speed up electronic structure calculations~\cite{lorenz2004representing, PhysRevLett.98.146401, balabin2009neural, handley2010potential, behler2011neural, schutt2017schnet, schutt2017quantum, xie2018permutation}, while fewer studies have been devoted to developing NN XC functionals~\cite{nagai2018neural, lei2019design, nagai2019completing, ramos2019static}. NN was firstly utilized to approximate XC potential by Tozer et al.~\cite{tozer1996exchange}. This study showed that NN approach is flexible and principally applicable to generate XC potentials to be used in the real physical systems DFT calculations. 

\begin{figure}
    \centering
    \includegraphics[scale=0.10]{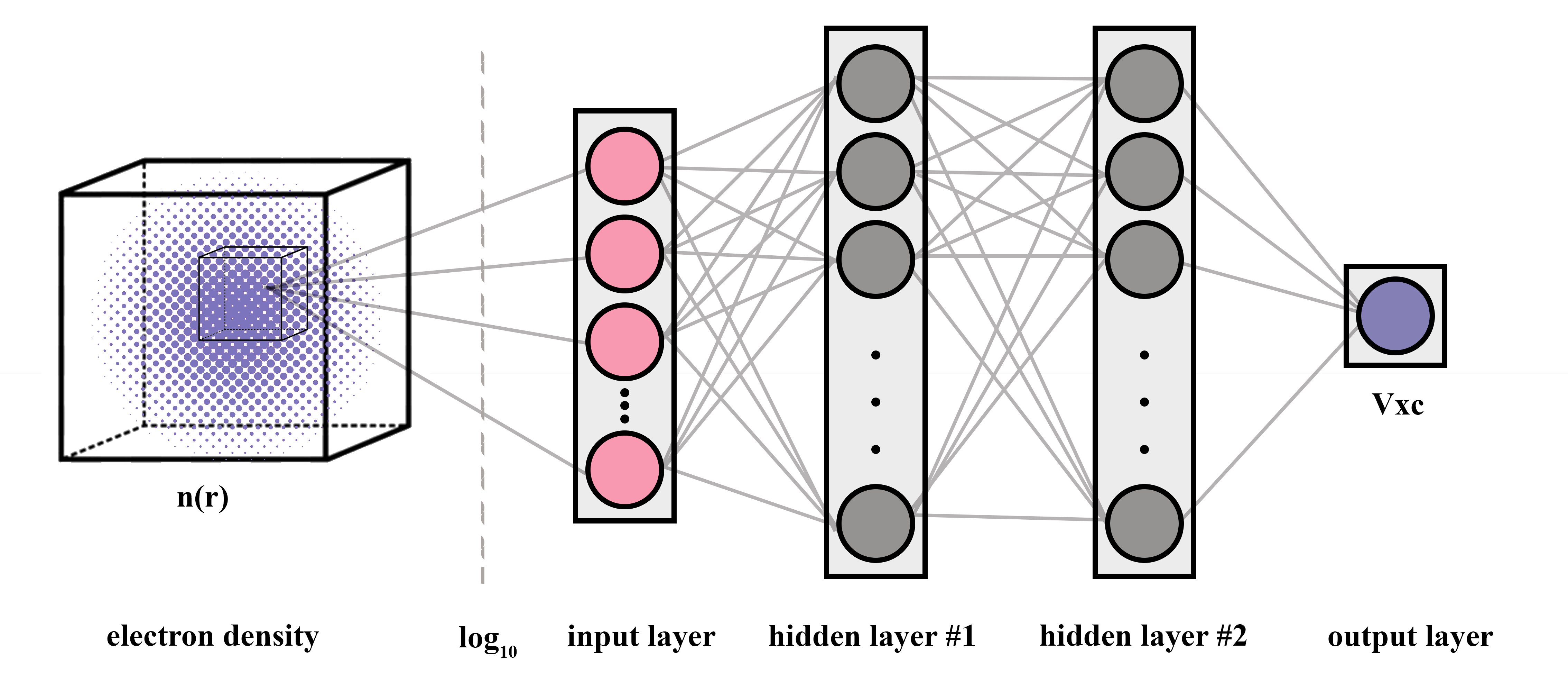}
    \caption{\textbf{Topology of the developed neural network. Two hidden layers are used, with 1000 and 500 neurons, respectively. A base 10 logarithm is applied to the input data (electron density). This helps to scale the input data to the range that the neural network is most sensitive to.}}
    \label{fig:nnstructure}
\end{figure}

In this work, we focus on constructing NN XC functionals for 3D systems without any feature engineering. We show that even rather simple NN architectures reproduce LDA and GGA XC potentials with satisfactory accuracy. To construct the LDA, potential point-to-point mapping is used due to the local idea behind the LDA itself. GGA potential region-to-point mapping is achieved by increasing the number of input neurons to take into account the local environment.

This paper is outlined as follows: In methods section, we describe the data generation process, NN topologies and  hyperparameters. In results section, we show that the constructed NN XC potentials have the capability to reproduce LDA and GGA results for real physical systems. In conclusion section, we summarize our results and consider future research that could be done with the proposed framework, with a particular emphasis on incorporating data from high-level methods such as quantum Monte Carlo and post-Hartree-Fock. 

\section*{Methods}

To obtain the training data needed to build the NN XC potential, we carry out calculations of three-dimensional (3D) electron gases in a framework of the Kohn-Sham formalism~\cite{hohenberg1964inhomogeneous, kohn1965self} for potentials of a simple harmonic oscillator (SHO). This potential takes the form:

\begin{equation}\label{eq:shoform}
V_{ext}(\textbf{r}) = \frac{1}{2}\sum \limits_{i}^{3} k_i x_i^2 
\end{equation}

\noindent
where $k_i$ is the spring constant and $\textbf{r} = (x_1, x_2, x_3)$ is the radius vector. Parameter $k_i$ pertains to a simple harmonic oscillator randomly sampled from intervals $0.01 \leq k_i \leq 0.9~\textrm{Ha}/\textrm{a}_0^2$, where $a_0$ is 1 Bohr radius.

\begin{table}[b]
    \centering
    \begin{tabular}{lcc}
        \multicolumn{1}{c}{}&
        \multicolumn{1}{c}{\texttt{NN LDA}}&
        \multicolumn{1}{c}{\texttt{NN GGA}}\cr
        \hline
        Num. of inputs& 1& 125\\
        Dataset size, $10^6$& 3.3& 3.3\\
        $\mathrm{r_s}$ range, Bohr& [1; $5 \times 10^6$] & [1; $1 \times 10^8$]\\
        Training / test split (\%) & 85:15 & 85:15
    \end{tabular}
    \caption{\textbf{General information about NN XC functionals training process}}
    \label{tab:nnoverview}
\end{table}

The XC potential is evaluated on a $40 \times 40 \times 40\ a_0$ parallelepiped with $32 \times 32 \times 32$ mesh, which corresponds to a spacing of $1.25~a_0$. Three electrons are placed in this 3D space. For each chosen parameter, DFT calculations in real space were performed using the Octopus code~\cite{andrade2015real, andrade2012time, andrade2013real}. The parametrizations of the LDA exchange-correlation functional used include the Slater exchange \cite{dirac_1930, bloch1929bemerkung} and Perdew-Zunger (modified) correlation part~\cite{PhysRevB.23.5048}. The GGA parameters used include the exchange and correlation parts from~\cite{perdew1996generalized, PhysRevLett.78.1396}. The total number of DFT calculations performed is 150 for each XC parametrization. The ranges of electron density represented by the Wigner-Seitz radius $r_s$ are $[1; 5 \times 10^6]$ and $[1; 10^8]$ for LDA and GGA, respectively. One calculation then provides a $32 \times 32 \times 32$ dataset size for LDA and a $28 \times 28 \times 28$ size for GGA. These yield overall dataset sizes of approximately 5.0 million and 3.3 million for LDA and GGA, respectively, but only 3.3 million samples are used for both functionals. 15\% of all datasets are used for the test set.

The composition of Tensorflow~\cite{abadi2016tensorflow} and TFLearn~\cite{tang2016tf} is utilized for training the NN. The Adam algorithm with a learning rate of 0.001 is used for training, with MSE loss. The neural architecture used for the LDA is a fully connected network with one input neuron, one linear output neuron and two ReLU~\cite{nair2010rectified} hidden layers with 1000 and 500 neurons (see fig.~\ref{fig:nnstructure}). The only difference in the GGA case is that it takes into account the contribution of the local environment, which corresponds to an increase in the number of neurons in the first layer to 125 ($5 \times 5 \times 5$ cube). It should be noted that in principle the local environment could be taken into account explicitly by passing gradient and higher order derivatives of electron density. We delegate a computation of electron density derivatives to NN itself, making preprocessing step easy and convenient. Considered $5 \times 5 \times 5$ input cube gives information to calculate gradient and Jacobian matrix with $O(\Delta r^2)$ and $O(\Delta r)$ precision correspondingly, where $\Delta r$ is mesh spacing.

\begin{figure*}
    \centering
    \includegraphics[trim={2cm 0 2cm 1cm}, clip, scale=0.55]{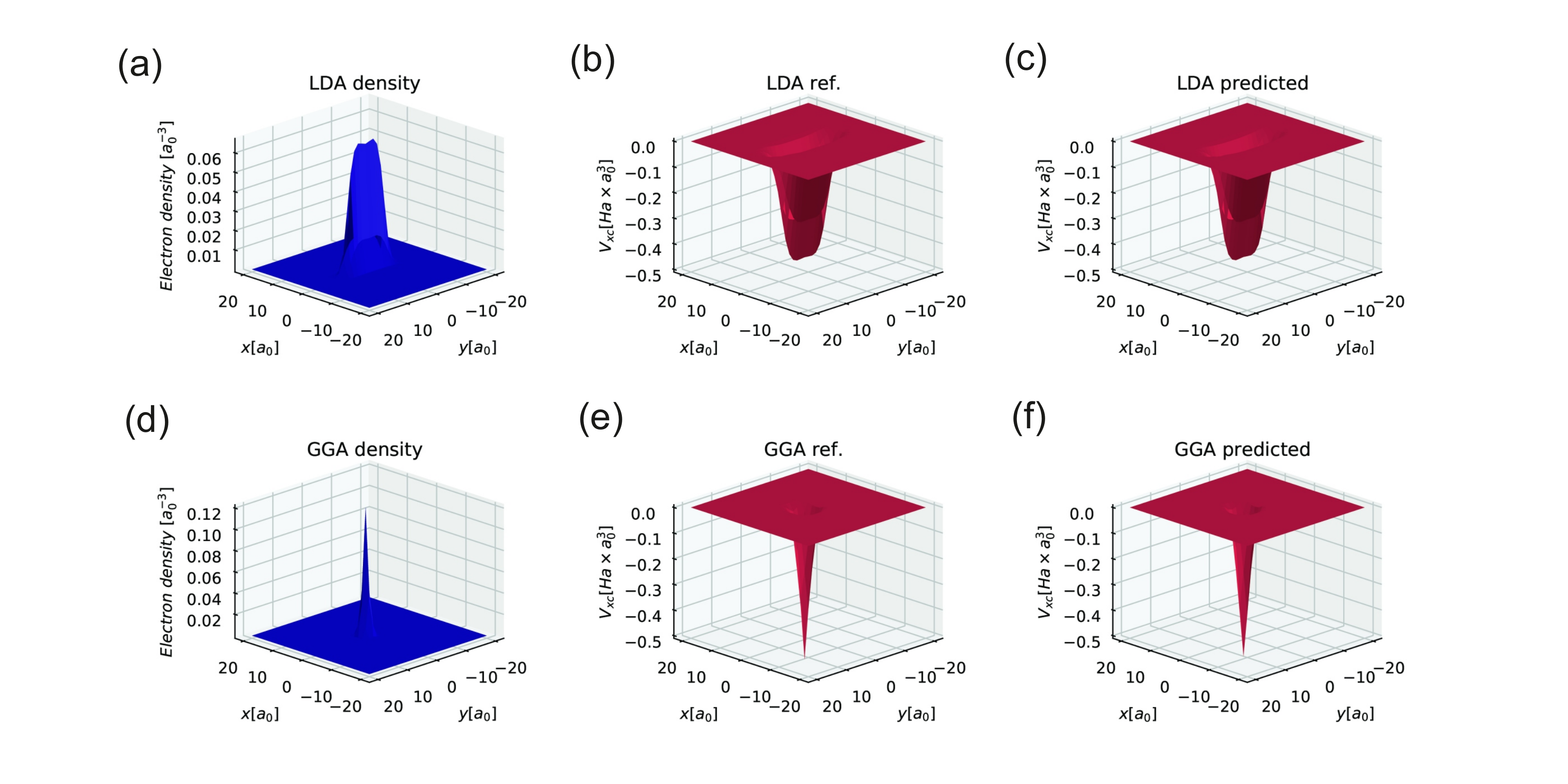}
    \caption{\textbf{The results of applying NN XC functionals to the self-consistent electron density obtained from solving the Schrodinger equation in the framework of DFT for 3 electrons in a simple harmonic potential. LDA and GGA "true" $V_{xc}$ results are produced by the Octopus package. All presented images are slices of 3D data in the z = 0 plane; (a) - the slice of the electron density obtained in the DFT calculation using LDA XC funtional, (b) - the slice of the XC potential obtained in the same as (a) DFT calculation using LDA XC functional, (c) - the slice of the NN XC potential obtained from the (a) electron density, (d) - the slice of the electron density obtained in the DFT calculation using GGA XC funtional, (e) - the slice of the XC potential obtained in the same as (d) DFT calculation using GGA XC functional, (f) - the slice of the NN XC potential obtained from the (d) electron density.}}
    \label{fig:synthetic} 
\end{figure*}

\subsection*{Data Availability}
Optimized NN weights and code to use it are available at \url{https://github.com/AlexanderFreeman/MLXC}.

\section*{Results and Discussion}

Multiple neural architectures, including different activation functions, numbers of layers and numbers of neurons, are tested. We choose robust and rather simple architectures, namely, the point-to-point and region-to-point architectures described above. We also test NNs with only one hidden layer with 1000 neurons, which have poor generalizability and result in a larger loss in comparison with the topology used. It should be noted that the topology chosen is heuristics, and it is possible that another number of hidden layers and neurons could give better results than those provided in this study. The search for the optimal NN topology is a large task that needs to be carried out in future studies.

Because the density range contains mostly small values, base-10 logarithmic transformation is chosen for preprocessing the dataset. This scaling greatly improves the convergence speed of the backpropagation algorithm during the NN training cycle. No other preprocessing techniques and feature engineering are applied to the input data. 

\begin{figure*}
    \centering
    \includegraphics[trim={2cm 0 2cm 1cm}, clip, scale=0.55]{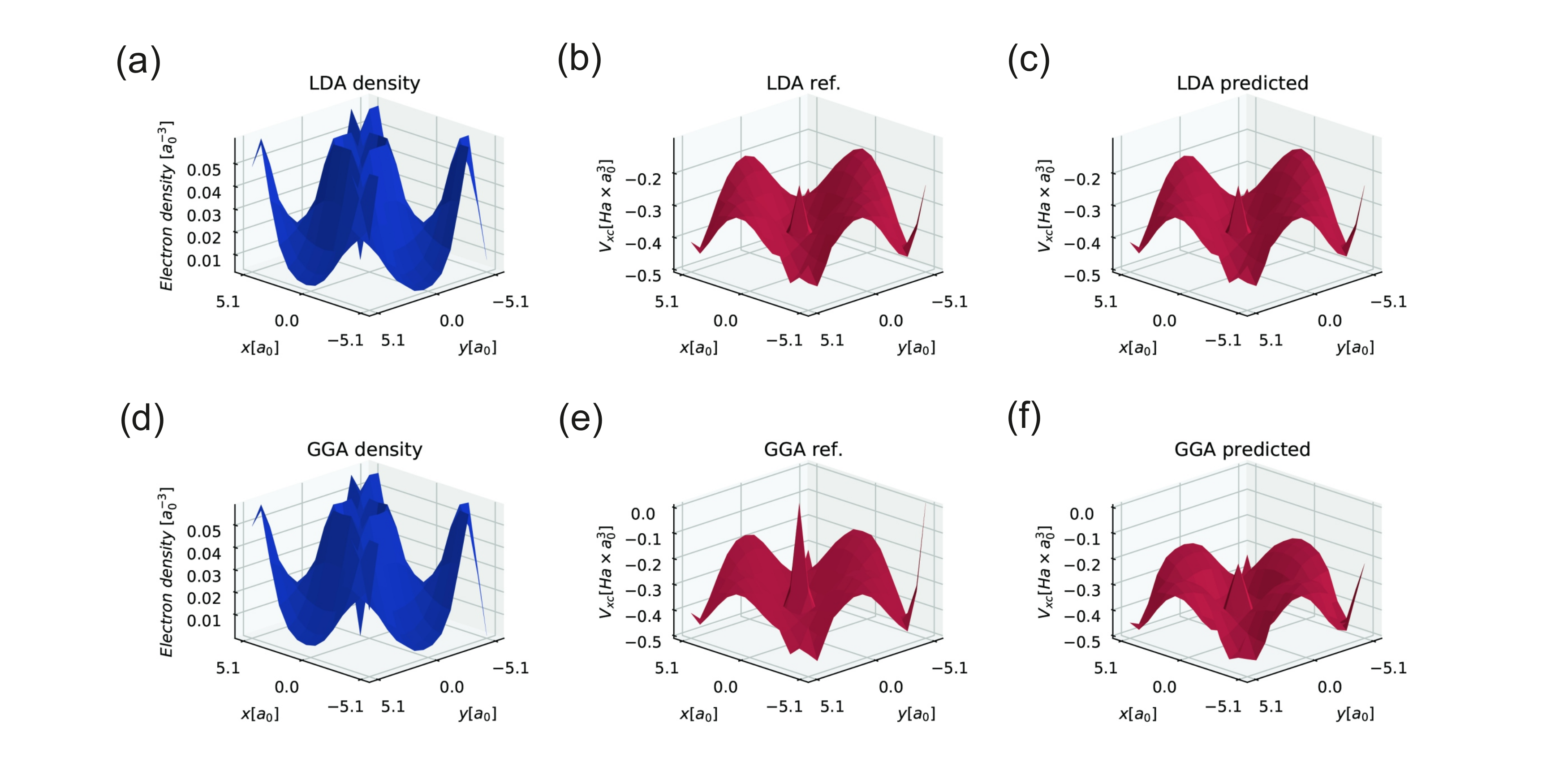}
    \caption{\textbf{The results of applying NN XC functionals to the self-consistent electron density obtained from solving the Schrodinger equation in the framework of DFT for a Si diamond structure with 8 basis atoms in a cubic cell. The LDA and GGA "true" $V_{xc}$ results are produced by the Octopus package. All presented images are slices of 3D data in the (001) plane defined by the Miller indices; (a) - the slice of the electron density obtained in the DFT calculation using LDA XC funtional, (b) - the slice of the XC potential obtained in the same as (a) DFT calculation using LDA XC functional, (c) - the slice of the NN XC potential obtained from the (a) electron density, (d) - the slice of the electron density obtained in the DFT calculation using GGA XC funtional, (e) - the slice of the XC potential obtained in the same as (d) DFT calculation using GGA XC functional, (f) - the slice of the NN XC potential obtained from the (d) electron density.}}
    \label{fig:silicon} 
\end{figure*}

Training of the LDA NN is stopped after 1300 epochs. The NN LDA mean absolute error (MAE) is 0.03~$\text{mHa} \times \text{Bohr}^3$. The mean minimum and maximum values of $V_{xc}$ averaged over all configurations used for the LDA NN construction are 0.73 and 510~$\text{mHa} \times \text{Bohr}^3$, respectively. Training of the GGA NN is terminated after 320 epochs. The NN GGA mean absolute error (MAE) is 0.156~$\text{mHa} \times \text{Bohr}^3$. The mean minimum and maximum values of Vxc averaged over all configurations used for the GGA NN construction are 0 and 553~$\text{mHa} \times \text{Bohr}^3$, respectively. The results of the training for both the LDA and GGA NNs are summarized in Table~\ref{tab:nnoverview}.

\begin{table}[b]
    \centering
    \begin{tabular}{lccc}
        \multicolumn{1}{c}{\texttt{}}&
        \multicolumn{1}{c}{\texttt{SHO}}&
        \multicolumn{1}{c}{\texttt{Si}}&
        \multicolumn{1}{c}{\texttt{Benzene}}\cr
        \hline
        Min Vxc, , $\text{mHa} \times \text{Bohr}^3$& 0.73& 187&4.69\\
        Max Vxc, $\text{mHa} \times \text{Bohr}^3$& 514& 495&755\\
        MAE, $\text{mHa} \times \text{Bohr}^3$& 0.03& 0.6& 1.36\\
        $n(\textbf{r})$ is in the training/test set& True& True& False\\
        Mesh is in the training/test set& True& False& False
    \end{tabular}
\caption{\textbf{Summary of results for NN LDA.}}
\label{tab:nnresultslda}
\end{table}

The results of applying an NN to the electron density from the test set are presented in fig.~\ref{fig:synthetic}. Both XC potentials provided by the LDA and GGA NNs are in good agreement with the numerical calculations made by Octopus. The special features of the $V_{xc}$ form, for example, the "two humps", are also well reproduced by them. For LDA it is also possible to provide comparison between $V_{xc}^{NN}(r_s)$ and analytical reference function $V_{xc}(r_s)$. Plotted curves are indistinguishable in the range of $r_s$ from 1 to 15, the absolute maximum difference is 1.7 $\text{mHa}\times \text{Bohr}^3$ (see fig.S1 of supplementary materials).

\begin{figure*}
    \centering
    \includegraphics[trim={2cm 0 2cm 1cm}, clip, scale=0.55]{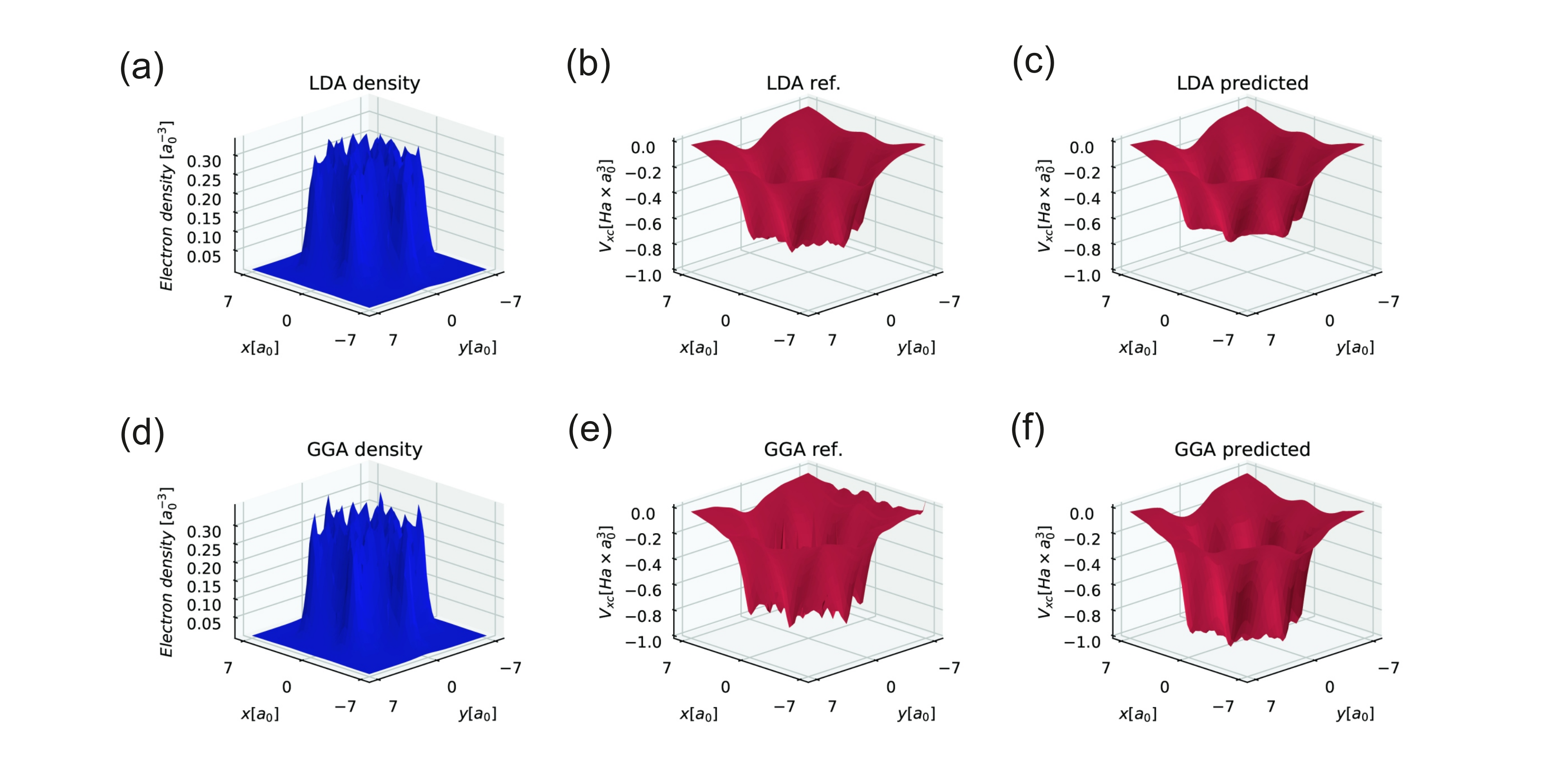}
    \caption{\textbf{The results of applying NN XC functionals to the self-consistent electron density obtained from solving the Schrodinger equation in the framework of DFT for a benzene molecule. The LDA and GGA "true" $V_{xc}$ results are procuced by the Octopus package. All presented images are slices of 3D data in the plane of the benzene molecule; (a) - the slice of the electron density obtained in the DFT calculation using LDA XC funtional, (b) - the slice of the XC potential obtained in the same as (a) DFT calculation using LDA XC functional, (c) - the slice of the NN XC potential obtained from the (a) electron density, (d) - the slice of the electron density obtained in the DFT calculation using GGA XC funtional, (e) - the slice of the XC potential obtained in the same as (d) DFT calculation using GGA XC functional, (f) - the slice of the NN XC potential obtained from the (d) electron density.}}
    \label{fig:benzene}
\end{figure*}

To test the NN XCs for real physical systems, we consider Si in a diamond structure with 8 basis atoms in a cubic cell (lattice constant: $10.2~a_0$). The numerical calculation is performed on a $10.2 \times 10.2 \times 10.2~a_0$ parallelepiped with a $14 \times 14 \times 14$ mesh, which corresponds to a spacing of $0.729~a_0$. The NN LDA and NN GGA MAEs are 0.6 and 18.1 ~$\text{mHa}\times \text{Bohr}^3$, respectively. The NN LDA results are in excellent agreement with the reference data obtained from Octopus. The NN GGA describes $V_{xc}$ reasonably well, except for regions with high density variations (see~\ref{fig:silicon}). We believe that this is because of the different spacings of the training/test data, which lead to the change in electron density gradients perceived by the NN XC.

\begin{table}[b]
    \centering
    \begin{tabular}{lccc}
        \multicolumn{1}{c}{\texttt{}}&
        \multicolumn{1}{c}{\texttt{SHO}}&
        \multicolumn{1}{c}{\texttt{Si}}&
        \multicolumn{1}{c}{\texttt{Benzene}}\cr
        \hline
        Min Vxc, , $\text{mHa} \times \text{Bohr}^3$& 0& 17&0.06\\
        Max Vxc, $\text{mHa} \times \text{Bohr}^3$& 553& 496&804\\
        MAE, $\text{mHa} \times \text{Bohr}^3$& 0.16& 18.1& 18.3\\
        $n(\textbf{r})$ is in the training/test set& True& True& False\\
        Mesh is in the training/test set& True& False& False
    \end{tabular}
    \caption{\textbf{Summary of training results for NN GGA}}
    \label{tab:nnresultsgga}
\end{table}

The NN XC potentials are also tested on slices of 3D data in the plane of the benzene molecule. The numerical calculation is performed on a $14 \times 14 \times 14\ a_0$ parallelepiped with a $44 \times 44 \times 44$ mesh, which corresponds to a spacing of $0.318~a_0$. The NN LDA and NN GGA MAEs are 1.36 and 18.3~$\text{mHa}\times \text{Bohr}^3$, respectively. The overall form of the XC potential obtained by Octopus is similar to the form produced by both NNs. In the case of benzene, the MAE is substantially larger than the MAE obtained during the training/test procedure. This is due to the high electron density values obtained for benzene, which are not presented in the training/test dataset. Moreover, the mesh spacing in the benzene calculation is smaller than that in the dataset, which is important for NN GGA potential evaluation. Despite the lack of appropriate input data, both NN XC potentials show an impressive ability to generalize the XC functional (see~\ref{fig:benzene}).

The results obtained from the developed NN XC potentials show that NN LDA works well, provided that the electron density values are presented in the training/test dataset. Based on table~\ref{tab:nnresultslda}, the NN LDA MAE is rather small for silicon and becomes larger for benzene, in which the electron density contains values that are outside the training/test density range. It is not surprising that the NN GGA (see table~\ref{tab:nnresultsgga}) demonstrates a large MAE for both systems considered, as the mesh spacing is not presented in the training procedure. Electron density spatial information not included explicitly in the training set leads to the fact that neural networks can take gradients into account in a way that cannot be properly generalized. Changing the mesh corrupts the learned representations and makes it not quite correct to apply the NN GGA to systems on meshes other than the one learned.

To construct mesh-independent nonlocal functionals such as the GGA, it is necessary to use the coordinates of mesh points at which the electron density is defined. This additional information will help the NN evaluate mesh-independent gradients or other higher-order variations of electron density and include them in the process of the interpolation of the exchange-correlation potential. The type of local coordinate system to use remains the subject of further research. 

\section*{Conclusions}
In this work, we developed, verified and tested a neural-network approach to interpolating XC functionals. We show that the NN approach gives adequate results for both the LDA and GGA considered. When considering a model system based on three electrons placed in a parallelepiped box with an external potential of a 3D simple harmonic oscillator, the NN is in perfect agreement with numerical calculations from the test dataset. Moreover, the neural network shows a satisfactory error in real 3D physical systems not presented in the training/test data, such as silicon and benzene. Despite the lack of generalizability of the NN GGA for both systems, the analysis of those results led us to the idea of improving the NN approach by searching for a better type of coordinate system for proper representation of the input data. It is important to note that the neural network topology used in this work is very simple and has good potential for further improvements.

The main advantage of the NN approach in comparison with other interpolation techniques for XC functionals is its flexibility to incorporate exchange-correlation data from different sources, such as post-Hartree-Fock and quantum Monte Carlo. It is possible that application of the NN to interpolate high-level XC quantum data could eliminate many heuristics used in the traditional construction of XC functionals.

\bibliography{sample}

\section*{Author contributions statement}
A.R. developed, trained and tested NN. A.R. and P.Z. analyzed the results. A.R. and P.Z. wrote the paper. P.Z and I.A. supervised the project. All authors reviewed the manuscript.

\section*{Additional information}

The authors declare that they have no competing interests.

\end{document}